\documentclass[%
 reprint,
 amsmath,amssymb,
 aps,
]{revtex4-2}
\usepackage{comment}
\usepackage{graphicx}
\graphicspath{figuresFolder/}
\usepackage{dcolumn}
\usepackage{bm}

\begin{document}

\preprint{APS/123-QED}

\title{Direct and retrograde signal propagation in unidirectionally coupled Wilson-Cowan oscillators}


\author{Guy Elisha$^{1*}$, Richard Gast$^{2*}$, Sourav Halder$^3$, Sara A. Solla$^2$, Peter J. Kahrilas$^{3,4}$, John E. Pandolfino$^{3,4}$, Neelesh A. Patankar$^{1+}$}

\affiliation{$^1$Department of Mechanical Engineering, Northwestern University}
\affiliation{$^2$Department of Neuroscience, Feinberg School of Medicine, Northwestern University}
\affiliation{$^3$Division of Gastroenterology and Hepatology, Feinberg School of Medicine, Northwestern University}
\affiliation{$^4$Kenneth C. Griffin Esophageal Center, Feinberg School of Medicine, Northwestern University}

\thanks{$^*$These authors contributed equally to this work}
\thanks{$^+$Corresponding author: N.~A.~Patankar (\texttt{n-patankar@northwestern.edu})}

\date{\today}

\begin{abstract}
\noindent 
Certain biological systems exhibit both direct and retrograde propagating wave signals, despite unidirectional neural coupling. However, there is no model to explain this. Therefore, the underlying physics of reversing the signal’s direction for one-way coupling remains unclear. Here, we resolve this issue using a Wilson-Cowan oscillators network. By analyzing the limit cycle period of various coupling configurations, we determine that intrinsic frequency differences among oscillators control wave directionality.

\end{abstract}

\maketitle


\section{\label{Introduction}Introduction}
Over the years, numerous studies focused on synchronization, directionality, and phase locking of coupled relaxation oscillators due to their ability to adequately describe many natural processes \cite{ashwin2016mathematical,schwemmer2012theory,strogatz1993coupled,strumillo2006application,cohen1982nature}. 
Finite chains of coupled nonlinear oscillators are used to model crawling, swimming, and peristalsis in biological organisms \cite{Pehlevan2016,Trayanova2011,daniel1994relaxation,collins1993coupled,ji2021phase,rowat1993modeling}. 
A chain of coupled uniform oscillators tends to produce a phase lag between adjacent oscillators, creating a propagating wave \cite{kopell2003chains,kopell1986symmetry}. 
The phase (size and direction) is a key component in many fields. For instance, it controls the direction and speed of muscle contractions in organs \cite{kirby1989coupling,Du2010,Pullan2004}, cognitive processes \cite{varela2001brainweb,brette2012computing,carlos2020anticipated}, and locomotion \cite{ijspeert2008central,matsushima1992neural,friesen1993mechanisms}. 

Switching between direct (negative phase) and retrograde (positive phase) propagating signal is an integral part of different systems \cite{matias2014modeling,hutchison2013dynamic}, species \cite{stein1971intersegmental,hughes2007sensory,haspel2010motoneurons}, and organ functions \cite{castedal1998postprandial,gharibans2016high}. However, it can also indicate a dysfunction in organs such as uterus and esophagus. In gravid uterus, unpredicted fundus to cervix contractions controlled by neural signals may indicate premature birth \cite{lange2014velocity}. In the esophagus, repetitive retrograde contractions observed during sustained volumetric distension of the organ suggests a motility disorder \cite{Carlson2020}. Revealing the mechanisms that control wave traveling directions in coupled oscillator systems is thus important for understanding the function of biological systems \cite{Carlson2020,bick2020understanding}.


Coupled oscillators models capable of producing both direct and retrograde propagating signals have been proposed in the past \cite{Gjorgjieva2013,sarna1972}. \citet{Gjorgjieva2013}, for example, presented a bidirectionally coupled network of oscillators. In their work, a direct wave was obtained through applying a fixed input (stimulating) only to the posterior oscillator, and a retrograde wave was obtained through applying a fixed input only to the anterior oscillator. The rest of the oscillators in the network did not receive an input and thus did not have a limit cycle solution by themselves. Switching the signal’s traveling direction in this network therefore requires relocation of the input.



In many biological settings, a network of unidirectionally coupled limit cycle oscillators is more appropriate \cite{cohen1992modelling,collins1993coupled,izhikevich2000phase,skinner1998intersegmental,ijspeert2008central,ryzhii2013modeling}. 
This differs from the previous study discussed, where the input was applied solely to either a distal or proximal oscillator, and the coupling was bidirectional \cite{Gjorgjieva2013}. 
For example, sustained volumetric esophageal distension triggers stretch receptors along the distended region. These receptors are simultaneously triggered, stimulating local excitatory neurons. This uniform input can be modeled as a unidirectionally coupled network of limit cycle relaxation oscillator with a fixed input, applied to all oscillators in the chain \cite{Omari2022,elisha2024neurological}.

Previous studies on chains of identical limit cycle oscillators suggest that transition in signal propagation direction occurs through changes in local frequency \cite{kopell1988coupled,cohen1992modelling}. This has been explored though multiple ventures such as implementing frequency gradients, applying external perturbations, and individually changing the inputs to each oscillator \cite{borisyuk2023phase,danner2016central,roxin2011effective,finger2019probing}. Varying the coupling weights and orientation have also shown to influence directionality \cite{kopell1988coupled,kopell1986symmetry,kopell2003chains}. 
However, for biological systems to implement any of the above mechanisms for switching the wave direction, it would place strong demands on the self-organization properties of the system, and thus make it prone to failure/pathology. Changing the synaptic coupling of a neural population, for example, is a slow process in comparison to the speed of traveling waves in the brain, rendering dynamic coupling adjustments of that match the timescale of behavioral demands challenging \cite{odonnell_nonlinear_2023}. 


Moreover, the majority of studies discussed above employed bidirectional coupling. Only a limited number of prior work focused on modeling unidirectional coupling \cite{kopell1990phase,borisyuk2023phase,kopell1988coupled}. Among these, only a select few successfully demonstrated both direct and retrograde propagating waves \cite{kopell1990phase,borisyuk2023phase}. However, despite these findings, none of the studies offered a comprehensive explanation for this phenomenon. Thus, the underlying physics remains unclear. 



Therefore, in this study, we aim to clarify the process by which wave direction reverses in a network of unidirectionally coupled limit cycle oscillaotrs. Additionally, we explore a mechanism that can easily and dynamically emerge in biological systems to obtain such transition. 
The directionality of the traveling wave can be reversed by changes in the global extrinsic drive to the system, a mechanism that requires much less self-organization for a biological system to implement than previously proposed mechanisms.

\section{\label{Formulation}Neural Model: Coupled Wilson-Cowan Oscillators}
\begin{figure}[b]
\includegraphics[trim=0 0 0 0,clip,width=.5\textwidth]{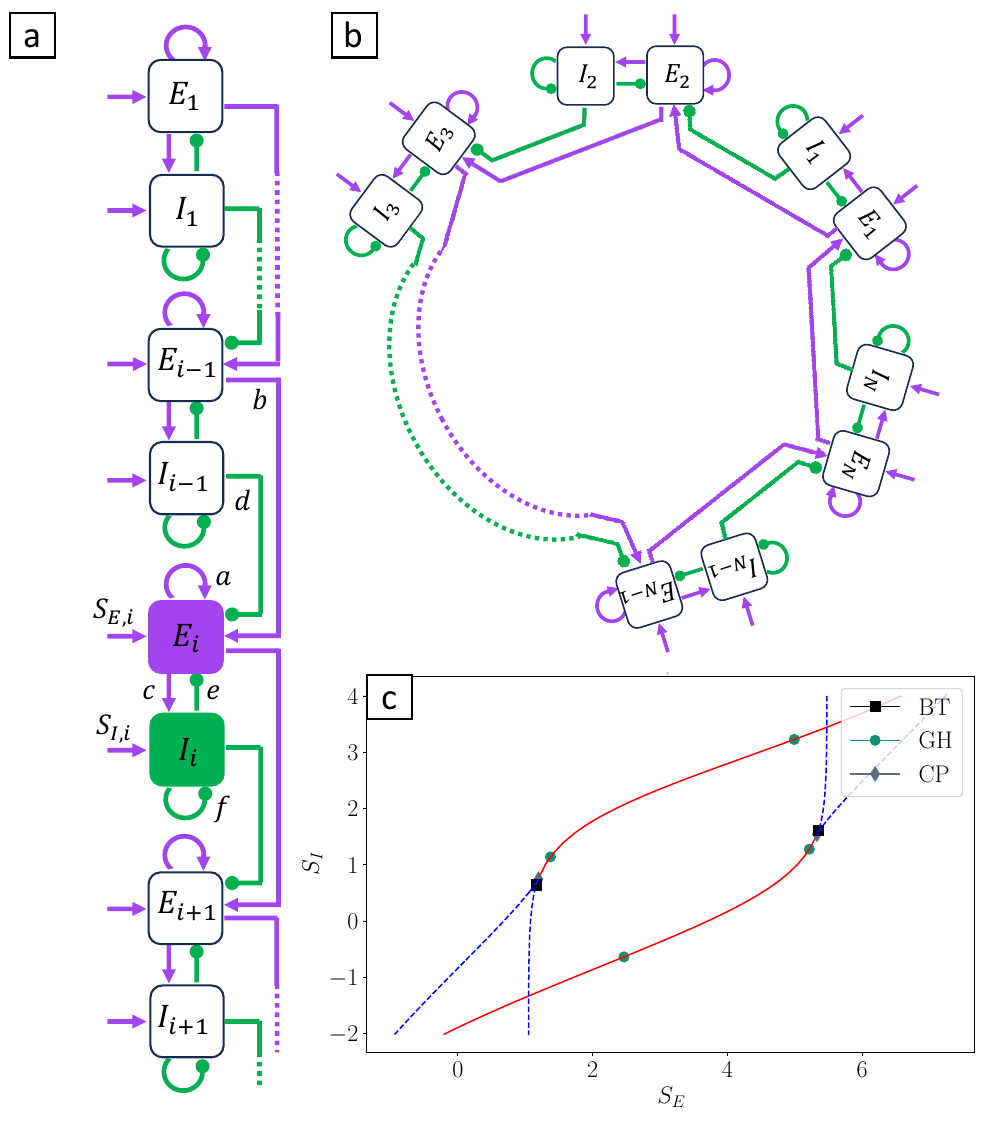}
\caption{\label{fig:schematic} Schematic of the model and the corresponding phase diagram. The neural network is composed of $N$ unidirectionally coupled relaxation oscillators, each consisting of excitatory ($E$) and inhibitory ($I$) neural populations. Purple lines mark excitatory synapses, and green lines denote inhibitory synapses. (a) Schematic of the coupled chain with the neuronal connections marked on segment $i$ ($a$, $b$, $c$, $d$, $e$, and $f$). The external input parameters are noted as $S_{E}$ and $S_{I}$. (b) Schematic of the ring configuration. (c) Phase diagram of oscillator 1 (decoupled) in the chain configuration. For the phase diagram of an oscillator in the ring configuration and bifurcation diagrams see \cite{supplement}. BT - Bogdanov-Takens bifurcation, GH - generalized Hopf bifurcation, CP - cusp bifurcation, dashed blue lines - fold bifurcation curves, solid red lines - Hopf bifurcation curves.}
\end{figure}

The neural model is composed of a chain of unidirectionally coupled Wilson-Cowan (WC) oscillators \cite{Wilson1972,Gjorgjieva2013} of the form
\begin{equation} \label{eq:Exc}
    \tau_{E}\dot{E}_i=-E_i+(1-E_i)\sigma_{E}[aE_i+bE_{i-1}-eI_i-dI_{i-1}+S_{E}],
\end{equation}
and
\begin{equation} \label{eq:Inh}
\tau_{I}\dot{I}_i=-I_i+(1-I_i)\sigma_{I}[cE_i-fI_i+S_{I}]
\end{equation}
(see schematic in Fig. \ref{fig:schematic}a). The WC model represents the evolution of excitatory ($E$) and inhibitory ($I$) activity in a synaptically coupled neuronal network. Subscript $i$ denotes oscillator $i$, where $i=1,...,N$. We set the number of oscillators to $N=70$. We define $i=1$ and $i=N$ as the proximal and distal ends, respectively. 

In the above equations, $\tau_{E}$ and $\tau_{I}$ are time constants, and the parameters $a$-$f$ symbolize the average number of synapses per cell in the respective population. 
Synaptic coupling between different WC oscillators is realized via $b$ and $d$, which represent the number of synapses formed by the excitatory and inhibitory populations of a WC oscillator with the excitatory population of the subsequent WC oscillator in the chain \cite{Gjorgjieva2013}.
The first WC oscillator receives no input from the rest of the network in the chain configuration, i.e. $b=d=0$ at $i=1$.
This is changed for the periodic boundary condition, also known as a closed chain or ring, where the first oscillator ($i=1$) receives input from the last oscillator ($i=N$) (Fig. \ref{fig:schematic}b). Here $b$ and $d$ are no longer equal to zero and Eq. (\ref{eq:Exc}) for $i=1$ becomes
\begin{equation} \label{eq:ExcRing}
    \tau_{E}\dot{E}_1=-E_1+(1-E_1)\sigma_{E}[aE_1+bE_{N}-eI_1-dI_{N}+S_{E}].
\end{equation}

The activation functions $\sigma_E$ and $\sigma_I$ are chosen as sigmoid functions of the form \cite{Wilson1972}
\begin{equation} \label{eq:sigmoid}
\begin{split}
\sigma_{E/I}[x] &= \frac{1}{1+\text{exp}[-\lambda_{E/I}(x-\phi_{E/I})]} \\
&\quad - \frac{1}{1+\text{exp}(\lambda_{E/I}\phi_{E/I})},
\end{split}
\end{equation}
\noindent which represents the non-linearity of the firing rate measured in neural populations in response to an input current ramp \cite{robinson1997propagation}. $\lambda$ and $\phi$ are the activation speed and activation threshold, respectively. Finally, $S_{E}$ and $S_{I}$ are extrinsinc inputs globally applied to all excitatory and inhibitory populations, respectively. The default values of all parameters are provided in Tab. \ref{table:param} \cite{Wilson1972,Gjorgjieva2013,elisha2024neurological}.

To ensure that stable limit cycle solutions of Eq.~\eqref{eq:Exc} and Eq.~\eqref{eq:Inh} exist for all parameter values considered in this study, we perform an initial bifurcation analysis of a single WC oscillator of the form \eqref{eq:Exc} and \eqref{eq:Inh} with $b = d = 0$, using the parameter continuation software PyCoBi \cite{gast_pyrates_2023}. 
As can be seen in Fig. \ref{fig:schematic}c, the bifurcation structure of the system in the 2D parameter space spanned by $S_E$ and $S_I$ is organized by two Bogdanov-Takens bifurcations that give rise to fold and Hopf bifurcation curves \cite{kuznetsov_elements_2013}. 
In the vicinity of the Bogdanov-Takens bifurcation, the system dynamics are governed by a stable fixed point and an unstable saddle point, which collide in a fold bifurcation. Additionally, a limit cycle emerges from the stable fixed through an Andronov-Hopf bifurcation.
The existence of this limit cycle is bounded by the saddle point curves, as the limit cycle disappears through a saddle homoclinic bifurcation. 
Due to the symmetry of this scenario, a region of stable oscillations exists between the Bogdanov-Takens bifurcation points (see Fig.\ref{fig:schematic}c).
The period of these oscillations is highly sensitive to changes in the control parameter $S_E$, as it grows towards infinity when approaching the homoclinic bifurcation points \cite{maruyama_analysis_2014,gonchenko_leonid_2022}.
Below, we examine the properties of signal propagation in chains of Wilson-Cowan oscillators in the limit cycle regimes.
Specifically, we study the traveling direction of wave solutions in the chain as we systematically change the proximity to the homoclinic bifurcation points via adjustments of the global input.



\begin{table}[b]
\caption{\label{table:param}%
List of parameters and their values
}
\begin{ruledtabular}
\begin{tabular}{lcdr}
\textrm{Symbol}&
\textrm{Value}\\
\colrule
                    $a$&16\\ 
                    $b$&20 \\ 
                    $c$&12\\ 
                    $d$&40 \\    
                    $e$&15\\ 
                    $f$&3 \\ 
                     $S_E$&1.14 to 5.27 \footnote{The $S_E$ parameter takes values within a range defined by the phase diagram in Fig.\ref{fig:schematic}} \\ 
                     $S_I$&-1.31 to 2.46 \footnote{The $S_I$ parameter takes values within a range defined by the phase diagram in Fig.\ref{fig:schematic}}\\     
                    $\phi_E$&4\\ 
                    $\phi_I$&3.7 \\ 
                    $\lambda_E$&1.3\\ 
                    $\lambda_I$&2 \\ 
                    $\tau_E$&1 \\  
                    $\tau_I$&4 \\   
\end{tabular}
\end{ruledtabular}
\end{table}

\section{Results}\label{Results}

\begin{figure*}[!htb]
    \centering{{\includegraphics[trim=0 0 0 0 ,clip,width=0.95\textwidth]{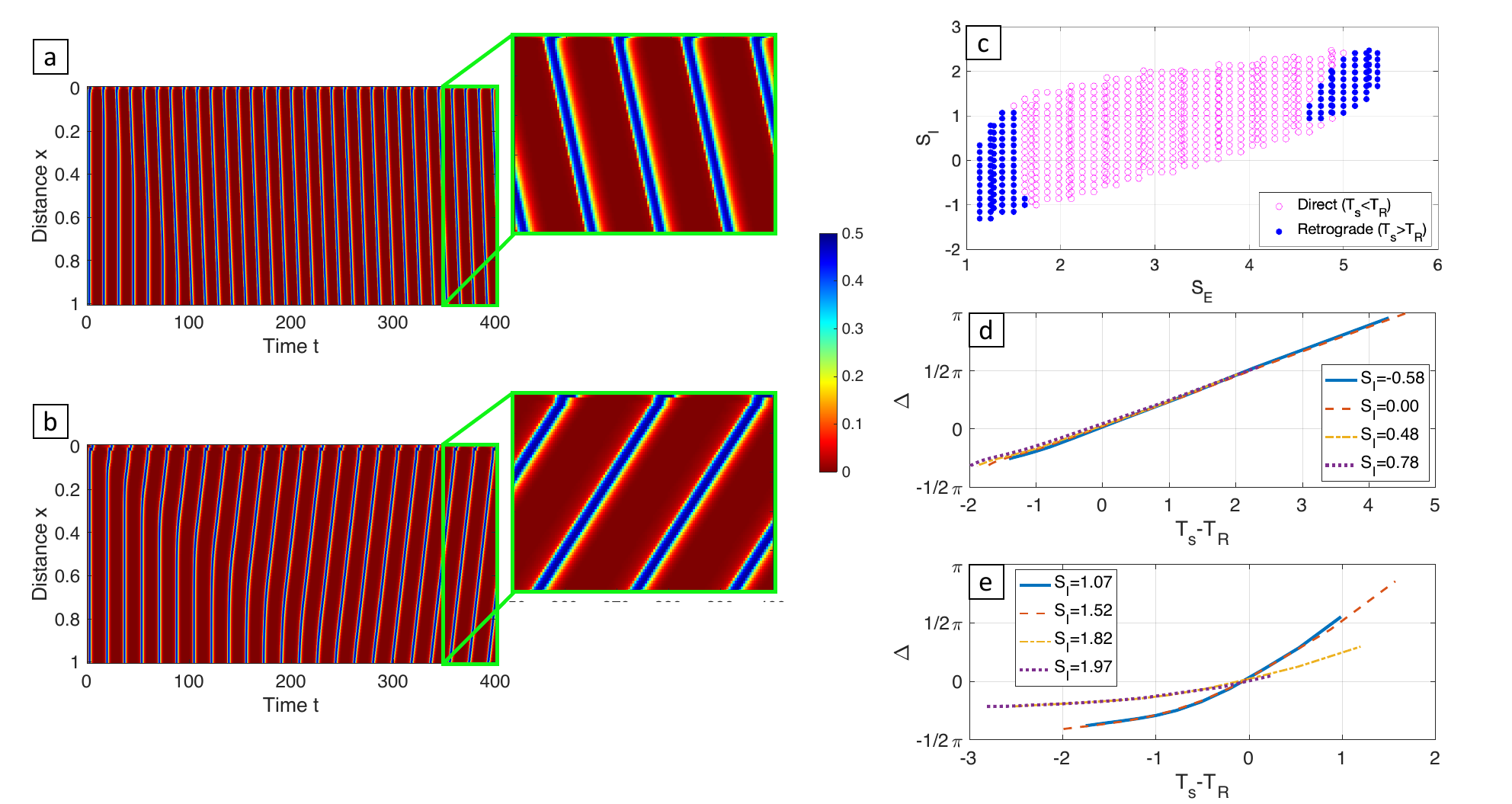}}}
    \caption{Spatio-temporal dynamics and phase transitions. (a) and (b) Spatio-temporal topography of the fast variable $E$ with $S_E = 2$ and $S_I=0$ (a), and $S_E = 1.4$ and $S_I=0$ (b). Parameter values are specified in Table \ref{table:param}. (c) Map used to identify propagating direction for a given chain based on the input value ($S_E$ and $S_I$). The classification is done based on the value of $T_s-T_R$. (d) and (e) Phase shift ($\Delta$) vs. $T_s-T_R$ curves for various $S_I$ values as $S_E$ changes. The phase is calculated over ten oscillators; values displayed represent ten times the actual phase between adjacent oscillators. In (d), $S_E$ values are less than 2.7, focusing on cases where decreasing $S_E$ transforms a direct wave into a retrograde wave. In (e), $S_E$ values are greater than 2.7, focusing on cases where increasing $S_E$ transforms a direct wave into a retrograde wave. This differentiation is motivated by the pattern observed in the phase diagram (Fig. \ref{fig:schematic}).}
    \label{fig:heatMaps}
\end{figure*}

We obtain our main results from numerical simulations of the WC chain network subject to systematic changes in the global input parameters $S_E$ and $S_I$.
For each combination of $S_E$ and $S_I$, we find that the network converges to a synchronized state, where all oscillators express phase locked dynamics at a global oscillation frequency that is identical to the decoupled, proximal WC oscillator frequency.
However, depending on the particular values of $S_E$ and $S_I$, the phases of the individual oscillators along the chain organize in stable patterns that correspond to either direct or retrograde wave propagation. 
Figure \ref{fig:heatMaps} displays the solution of the fast variable $E$ in the coupled chain with $S_I=0$ for two distinct levels of input to the excitatory cell populations ($S_E=2.0$ and $S_E=1.4$). As can be seen, decreasing $S_E$ results in a reversal of the wave propagation direction in the chain.

How can a change in global input parameters cause a reversal in the phase coupling pattern between oscillators along the chain?
To provide insights into this phenomenon, we examine whether similar behavior could be found in the ring configuration.
We find that all oscillators in the model oscillate at the exact same phase in the ring configuration scenario, independent of the choice of $S_E$ and $S_I$ within the range of considered values. 
Therefore, in Eq. (\ref{eq:Exc}), $a$ and $e$ become $a^*=a+b$ and $e^*=e+d$, which simplifies to
\begin{equation} \label{eq:ExcStar}
    \tau_{E}\dot{E}_i=-E_i+(1-E_i)\sigma_{E}[a^{*}E_i-e^{*}I_i+S_{E}].
\end{equation}

Most importantly, for the same global input values, the WC oscillators converge to a different oscillation frequency in the ring configuration than in the chain configuration.
Apparently, the additional network edge that is added to obtain the ring configuration is capable of both speeding up and slowing down the global oscillation frequency of the system (Fig.~\ref{fig:heatMaps}c).  
We note here again, that the only difference between the chain and ring configurations is the total synaptic input received by the first oscillator in the chain.
Thus, a difference must exist in the intrinsic oscillation frequency of the first oscillator in the chain configuration and any other oscillator further down the chain.
This frequency difference must correspond to the frequency differences between the ring configuration and the chain configuration shown in Fig.~\ref{fig:heatMaps}c.
We denote the period of oscillation of the first oscillator as $T_s$ in the chain configuration, and $T_R$ in the ring configuration.

In the chain configuration, the first WC oscillator ($i=1$) initiates the influence on its adjacent oscillator, forcing it to adjust its period to $T_s$. This influence gradually propagates down the chain, prompting distal oscillators to either decelerate or accelerate until their periods converge to $T_s$ \cite{supplement}. As a result, the oscillator at $i=1$ acts as the driving or forcing oscillator.
If chain oscillators are forced to speed up (i.e. $T_s<T_R$), a direct propagating wave emerges (see Fig.~\ref{fig:heatMaps}a). Conversely, if chain oscillators are forced to slow down (i.e. $T_s>T_R$), a retrograde propagating wave emerges (see Fig.~\ref{fig:heatMaps}b). The switch in propagating direction occurs when $T_s=T_R$ (Fig.~\ref{fig:heatMaps}d-e). For an illustration of tracking the period of the distal oscillator over time in multiple cases, refer to \cite{supplement}.

Interestingly, the figures demonstrate that adjusting the input to the excitatory population ($S_E$) is sufficient to reverse the direction of signal propagation. In fact, the value of $S_I$ hardly plays a role, as depicted in Fig. \ref{fig:heatMaps}c. The figure reveals that the $S_E$-$S_I$ space is mostly separated along the $S_E$ axis. Additionally, depending on the value of $S_E$, two scenarios can emerge. On the one hand, an increase in $S_E$ can transform a direct wave into a retrograde wave. On the other hand, a decrease in $S_E$ can change a direct wave into a retrograde wave.





\begin{figure}[!htb]
\includegraphics[trim=70 130 90 140,clip,width=.47\textwidth]{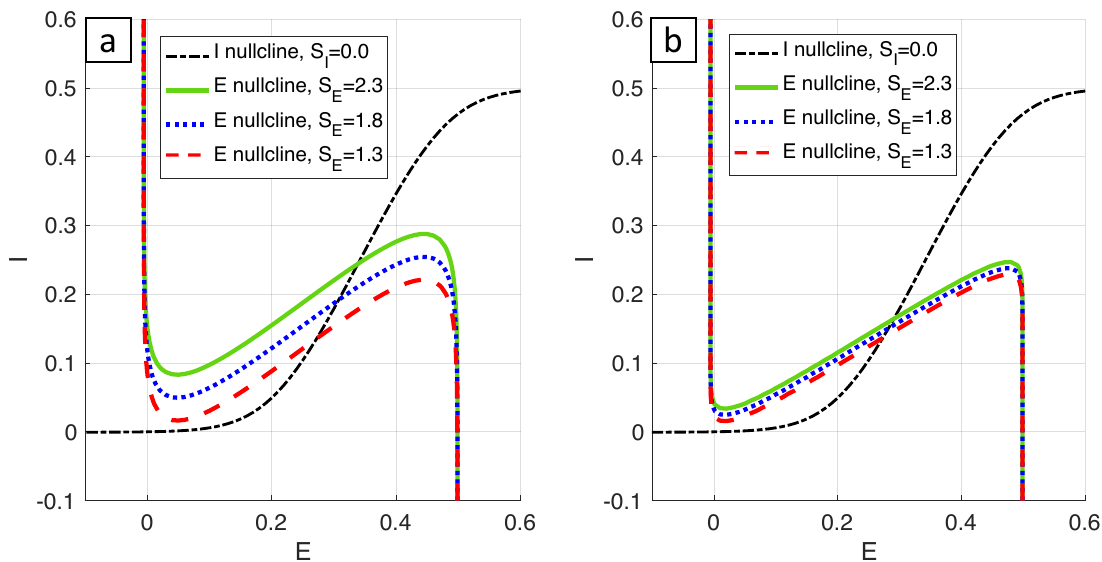}
\caption{\label{fig:phaseDiagrams} $E$ and $I$ nullclines of two different WC oscillators with various $S_E$ values. (a) Nullclines of a decoupled, single oscillator, defined by the parameter values listed in table \ref{table:param}. (b) Nullclines of an oscillator in the ring configuration. It is defined by the parameter values listed in table \ref{table:param}, but parameters $a$ and $e$ are set to the effective parameters, $a^*$ and $e^*$ from Eq. (\ref{eq:ExcStar}).
}
\end{figure}

While the excitatory ($S_E$) and inhibitory ($S_I$) inputs are uniform across all oscillators, their effects differ between the proximal oscillator ($i=1$) and the subsequent units due to coupling dynamics. A small change in $S_E$ can have substantial influence on the period of oscillation for the proximal oscillator, in contrast to the distal units, where intra-population innervations tend to dominate over external inputs (see Fig. \ref{fig:phaseDiagrams}). 
Since the proximal oscillator is the driving oscillator, the disproportionate impact of external inputs on its behavior plays a pivotal role in dictating the propagation direction.

Examining the $E$ and $I$ nullclines (Eq.(\ref{eq:Exc}) and Eq.(\ref{eq:Inh})) sheds light on this observation. In Fig. \ref{fig:phaseDiagrams}a, nullclines of a decoupled oscillator with period $T_s$ ($a$ and $e$ given in table \ref{table:param}) reveal that varying $S_E$ induces substantial shifts in the phase diagram. With increasing $S_E$ (1.3 to 2.3), the frequency of the limit cycle ($1/T_s$) rises, as the intersection between the two nullclines moves closer to the midpoint of the curves.

The nullclines in Fig. \ref{fig:phaseDiagrams}b, representing an oscillator in the ring configuration, with period $T_R$, illustrate a different trend. Varying $S_E$ leads to minor shifts in the phase diagram (Fig. \ref{fig:phaseDiagrams}b). In this scenario, intra-population innervation dominates, such that the frequency ($1/T_R$) barely changes as $S_E$ varies. Consequently, although increasing $S_E$ results in a simultaneous decrease in both $T_s$ and $T_R$, the reduction in $T_s$ is steeper, leading to an inversion from retrograde to direct signal propagation with increasing $S_E$. Refer to \cite{supplement} for comprehensive plots of $T_s$ and $T_R$ under diverse $S_E$ and $S_I$ conditions.



\section{Discussion}\label{Discussion}

The emergence of synchronized oscillations in neural systems and their spatial evolution as traveling waves has been shown to play a central role in neural system functions such as motor control and cognitive processes \cite{matias2014modeling,hutchison2013dynamic, stein1971intersegmental,hughes2007sensory,haspel2010motoneurons,castedal1998postprandial,gharibans2016high}.
In this context, dynamic adjustments of phase relationships between neural oscillators play a crucial role in switching between spatiotemporal synchronization patterns that are associated with different system functions \cite{hahn_portraits_2019,finger2019probing}.
Separate from this literature, neural systems have previously been proposed to operate close to a Bogdanov-Takens bifurcation, i.e. in a critical regime where small changes in the configuration of the neural system can trigger substantial changes in the system dynamics \cite{buendia2022broad,forrester2020role,gross2021not,ehsani2023scale}. 
However, how the vicinity to such a regime might affect the coordination of coupled neural oscillator systems has not been addressed so far. 

Here, we studied this scenario in a chain of coupled WC models - a neural oscillator model that expresses two Bogdanov-Takens bifurcations. 
Near those bifurcation points, transitions between oscillatory and steady-state dynamics occur via a saddle homoclinic bifurcation, where the limit cycle solution collides with a saddle point solution. 
In the vicinity of this bifurcation, neural oscillations are particularly sensitive to changes in the bifurcation parameter: relatively small changes in the input to WC model can lead to substantial changes in the period of oscillation. For a given level of global input, the initial chain oscillator and the other chain oscillators will be at different distances to one of the two homoclinic bifurcations they can express, thus causing substantial differences in their intrinsic frequencies. 
Since we have found that switching between direct and retrograde wave traveling directions in chains of coupled oscillators is controlled by the frequency differences between the initial chain oscillator and the other oscillators, this regime is particularly well suited for the control of wave traveling directions. For example, pushing the initial chain oscillator closer to the homoclinic bifurcation slows down its oscillation frequency considerably, and can thus cause a switch from direct to retrograde wave traveling in the oscillator chain.

While this study primarily focuses on theoretical implications for biological sciences, the concept of switching propagating signals may hold significant relevance in engineering applications. Coupled oscillators are often used to model central pattern generators for robotic locomotion \cite{ijspeert2008central,ramdya2023neuromechanics,drotman2021electronics}. Understanding how biological systems achieve efficient autonomous control is crucial for enhancing locomotor fluency in robots \cite{ramdya2023neuromechanics,lobato2022neuromechfly}. One of the pressing questions in this field involves modeling and executing gait transitions \cite{ijspeert2008central,ijspeert2007swimming}. Our findings introduce a sophisticated circuit (central pattern generator) capable of generating and switching between two different locomotor behaviors. The simplicity of obtaining switching through our proposed mechanism makes it a potential candidate for future exploration in robotics.

In conclusion, we have found that the vicinity to a Bogdanov-Takens bifurcation endows neural oscillator systems with an inherent mechanism to flexibly adjust properties of their spatiotemporal synchronization patterns such as the wave traveling direction. 
Near the Bogdanov-Takens bifurcation, local differences in the synaptic innervation across a neural oscillator system will translate to differences in the proximity to a homoclinic bifurcation. 
As long as these local changes persist, global changes in the level of excitation can reliably trigger changes in the phase configuration of the system, and thus in the wave traveling direction.
Since neuromodulators such as dopamine, serotonin, or acetylcholine can reliably cause such global, dynamic changes in the level of excitation \cite{lee_neuromodulation_2012}, we propose this as a self-organization mechanism of neural systems to flexibly react to changing behavioral demands, without relying on slow, plasticity-related changes of the synaptic coupling or oscillator properties.
Finally, the Bogdanov-Takens bifurcation is not specific to the WC model, but has been shown to exist in various neural population configurations \cite{gast_mean-field_2020,ehsani2023scale,gast_neural_2024}, thus emphasizing its importance for the dynamic configuration of neural systems.

\nocite{*}

\bibliographystyle{apsrev4-1}
\bibliography{apssamp}

\end{document}